\documentclass[a4paper]{jpconf}
\usepackage{graphicx}

\newcommand{\KV}{{\mbox{$\kappa \sigma^{2}$}}}
\newcommand{\SD}{{\mbox{$S \sigma$}}}

\newcommand{\gevc}{\mbox{${\mathrm{GeV/}}c$}}
\newcommand{\gev}{\mbox{$\mathrm{GeV}$}}
\newcommand{\sNN}{{{$\sqrt{s_{_{\mathrm{NN}}}}$}}}
\newcommand{\dNdeta}{\mbox{$dN_{ch}/d\eta$}}
\newcommand{ \be }{\begin{equation}}
\newcommand{ \ee }{\end{equation}}

\begin{document}
\title{Probing the QCD Critical Point with Higher Moments of Net-proton Multiplicity
Distributions}

\author{Xiaofeng Luo for the STAR Collaboration}

\address{Department of Modern Physics, University of Science and Technology of China, Hefei 230026, China}
\address{Nuclear Science Division, Lawrence Berkeley National Laboratory, Berkeley, CA 94720, USA}

\ead{xfluo@lbl.gov}

\begin{abstract}
Higher moments of event-by-event net-proton multiplicity
distributions are applied to search for the QCD critical point in
the heavy ion collisions. It has been demonstrated that higher
moments as well as moment products are sensitive to the correlation
length and directly connected to the thermodynamic susceptibilities
computed in the Lattice QCD and Hadron Resonance Gas (HRG) model. In
this paper, we will present measurements for kurtosis ($\kappa$),
skewness ($S$) and variance ($\sigma^2$) of net-proton multiplicity
distributions at the mid-rapidity ($|y|<0.5$) and $0.4<p_{T}<0.8$
{\gevc} for Au+Au collisions at {\sNN}=19.6, 39, 62.4, 130 and 200
GeV, Cu+Cu collisions at {\sNN}=22.4, 62.4 and 200 GeV, d+Au
collisions at {\sNN}=200 GeV and p+p collisions at {\sNN}=62.4 and
200 GeV. The moment products $\kappa \sigma^2$ and $S \sigma$ of
net-proton distributions, which are related to volume independent
baryon number susceptibility ratio, are compared to the Lattice QCD
and HRG model calculations. The {\KV} and {\SD} of net-proton
distributions are consistent with Lattice QCD and HRG model
calculations at high energy, which support the thermalization of the
colliding system. Deviations of $\kappa \sigma^2$ and $S\sigma$ for
the Au+Au collisions at low energies from HRG model calculations are
also observed.

\end{abstract}

\section{Introduction}
One of the main goals of heavy ion collision is to explore the phase
structure of hot, dense nuclear matter~\cite{starwhitepaper}. Finite
temperature Lattice QCD calculations demonstrate that with vanishing
baryon chemical potential ($\mu_B=0$), the phase transition from the
hadronic phase to the Quark Gluon Plasma (QGP) phase is a smooth
crossover~\cite{crossover}. The corresponding transition temperature
is about $170-190$ MeV~\cite{temperature,science}. At large $\mu_B$
region, the phase transition between hadronic phase and QGP phase is
of first order~\cite{firstorder} and with a second order end point
at the boundary towards the crossover region, which is so called QCD
Critical Point (CP)~\cite{qcp}. Although many efforts have been made
by theorist and experimentalist to search for the CP, its location
or even the existence is still not confirmed yet~\cite{location}.
The first principle Lattice QCD calculation at finite $\mu_B$ are
difficult due to the fermion sign problem. Several techniques, such
as Re-weighting, Image baryon chemical potential and Taylor
expansion~\cite{methods} $\it etc.$, have been developed to overcome
those problems and make the Lattice QCD calculable at finite $\mu_B$
region. However, large uncertainties are still there.

Experimentally, heavy ion collision provides us a good opportunity
to search for the CP. To access the QCD phase diagram, we can tune
the chemical freeze out temperature ($T$) and baryon chemical
potential ($\mu_B$) by varying the colliding energy. As the
characteristic signatures of the CP in a static and infinite medium
is the divergence of the correlation length ($\xi$) and increase of
non-Gaussian fluctuations. Thus, non-monotonic signals of CP are
expected to be observed if the evolution trajectory ( $T,\mu_{B}$ )
in the QCD phase diagram of the system pass nearby the critical
region and the signals are not washed out by the expansion of the
colliding system.

Due to the finite size effect, rapid expansion and critical slowing
down {\it etc.}, the typical correlation length ($\xi$) developed in
the heavy ion collision near the QCD critical point is a small value
about $2-3$ fm~\cite{correlationlength}. Recently, model
calculations reveal that higher moments of conserved quantities
distributions are proportional to the higher power of the
correlation length~\cite{qcp_signal,ratioCumulant}, such as fourth
order cumulant $<(\delta N)^{4}>-3<(\delta N)^{2}>^{2}\sim \xi^{7}$
, where $\delta N=N-M$, $N$ is the particle multiplicity in one
event and $M$ is the averaged particle multiplicity of the event
sample. On the other hand, the higher moments as well as moment
products of conserved quantities distributions are also directly
connected to the corresponding susceptibilities in Lattice QCD
~\cite{Lattice,MCheng2009} and HRG model~\cite{HRG} calculations,
for e.g. the third order susceptibility of baryon number
($\chi^{(3)}_{B}$) is related to the third cumulant ($<(\delta
N_{B})^{3}>$) of baryon number distributions as
$\chi^{(3)}_{B}={<(\delta N_{B})^{3}>}/{VT^3}$; $V,T$ are volume and
temperature of system respectively. It has been found that the
volume independence baryon number susceptibility ratio can be
directly connected to the moment products of baryon number
distributions as $\kappa \sigma^2=\chi^{(4)}_{B}$/$\chi^{(2)}_{B}$
and $S \sigma=\chi^{(3)}_{B}$/$\chi^{(2)}_{B}$, which allows us to
compare the theoretical calculations with experimental measurements.
Theoretical calculations also demonstrate that the experimental
measurable net-proton number (proton minus anti-proton number)
event-by-event fluctuations can reflect the baryon number and charge
fluctuations~\cite{Hatta}. Thus, higher moments of the net-proton
multiplicity distributions are applied to search for the QCD
critical point in the heavy ion
collisions~\cite{CPOD2010,PRL,SQM2009}. When approaching the QCD
critical point, the moment products {\KV} and {\SD} will show large
deviation from its Poisson statistical value. The skewness is
expected to change its sign when system evolution trajectory in the
phase diagram cross phase boundary~\cite{Asakawa}.

In year 2010, RHIC Bean Energy Scan (BES) program~\cite{bes} was
carried out to map the first order QCD phase boundary and search for
the QCD critical point by tuning the colliding energy from 39 GeV
down to 7.7 GeV with the corresponding $\mu_{B}$ coverage about
$100-410$ MeV. With the large uniform acceptance and good capability
of particle identification STAR detector, it provides us very good
opportunities to find the QCD critical point with sensitive
observable, if the existence of QCD critical point is true.

\section{Experimental Method}
The data presented in the paper are obtained using the Solenoidal
Tracker at RHIC (STAR). Those are Au+Au collisions at {\sNN}=19.6
(year 2001), 39, 62.4 (year 2004), 130 and 200 GeV (year 2004),
Cu+Cu collisions at {\sNN}=22.4, 62.4, 200 GeV, d+Au at {\sNN}=200
GeV (year 2003) and p+p collisions at {\sNN}=62.4 (year 2006), 200
GeV (year 2009). The main subsystem used in this analysis is a
large, uniform acceptance cylindrical Time Projection Chamber (TPC)
covering a pseudo-rapidity range of $|\eta|<1$ and $2\pi$ azimuthal
coverage. As a primary tracking device, it can measure the
trajectories and momenta of particles with transverse momenta above
$0.15$ {\gevc}. To ensure the purity and similar efficiency, the
protons and anti-protons are identified with the ionization energy
loss ($dE/dx$) measured by TPC of STAR detector within
$0.4<p_{T}<0.8$ {\gevc} and mid-rapidity ($|y|<0.5$). Several track
quality cuts are also used to select the tracks with good quality in
each event. We require the distance of closest approach (dca) to the
primary vertex of proton (antiproton) track less than 1 cm to
suppress the contamination from secondary protons (antiproton). To
study the centrality dependence of higher moments, centralities are
determined by the uncorrected charged particle multiplicities
($dN_{ch}/d\eta$) within pseudo-rapidity $|\eta|<0.5$ measured by
TPC. By comparing measured $dN_{ch}/d\eta$ with the Monto Carlo
Glauber model results, we can obtain the average number of
participant $N_{part}$, impact parameter $b$ and number of binary
nucleon-nucleon collisions $N_{bin}$ for each centrality class.

\section{Analysis Method}
In this section, we will introduce you how to perform the higher
moments analysis with experimental data. First, we will show you the
definition of the cumulants and various moments used in our
analysis, such as standard deviation ($\sigma$), skewness ($S$) and
kurtosis ($\kappa$). Then, we will discuss about the {\it Centrality
Bin Width Effect }(CBWE).
\subsection{Moments and Cumulants of Event-by-Event Fluctuations}
In statistics, probability distribution functions can be
characterized by the various moments, such as mean ($M$), variance
($\sigma^2$), skewness ($S$) and kurtosis ($\kappa$). Before
introducing the above moments used in our analysis, we would like to
define cumulants, which are alternative methods to the moments of a
distribution. The cumulants determine the moments in the sense that
any two probability distributions whose cumulants are identical will
have identical moments as well.

Experimentally, we measure net-proton number event-by-event wise,
$N_{p-\bar{p}}=N_{p}-N_{\bar{p}}$, which is proton number minus
antiproton number, in the mid-rapidity ($|y|<0.5$) and within the
transverse momentum $0.4<p_{T}<0.8$ {\gevc}. In the following, we
use $N$ to represent the net-proton number $N_{p-\bar{p}}$ in one
event. The average value over whole event ensemble is denoted by
$<N>$, where the single angle brackets are used to indicate ensemble
average of an event-by-event distributions.

The deviation of $N$ from its mean value are defined by
\begin{equation}
  \delta N=N-<N>.
\end{equation}
Then, we can define the various order cumulants of event-by-event
distributions as:
\begin{eqnarray}
C_{1,N}=<N>, C_{2,N}=<(\delta N)^{2}>, C_{3,N}=<(\delta N)^{3}>,
C_{4,N}=<(\delta N)^{4}>-3<(\delta N)^{2}>^{2}.
\end{eqnarray}

An important property of the cumulants is their additivity for
independent variables. If $X$ and $Y$ are two independent random
variables, then we have $C_{i,X+Y}=C_{i,X}+C_{i,Y}$ for $i$th order
cumulant. This property will be used in our study.

Once we have the definition of cumulants, various moments can be
denoted as:
\begin{eqnarray}
M=C_{1,N},\sigma^{2}=C_{2,N},S=\frac{C_{3,N}}{(C_{2,N})^{3/2}},\kappa=\frac{C_{4,N}}{(C_{2,N})^{2}}
\end{eqnarray}

And also, the moments product $\kappa \sigma^{2}$ and $ S \sigma$
can be expressed in term of cumulant ratio.

\begin{eqnarray}
\kappa \sigma^{2}=\frac{C_{4,N}}{C_{2,N}},
S\sigma=\frac{C_{3,N}}{C_{2,N}}.
\end{eqnarray}

With above definition of various moments, we can calculate various
moments and moment products with the measured event-by-event
net-proton fluctuations for each centrality.

\subsection{Centrality Bin Width Effect (CBWE) Correction}
The centralities in this analysis are determined by the uncorrected
charged particle multiplicity ($N_{ch}$) measured at middle
pseudo-rapidity ($|\eta|<0.5$) by the TPC of the STAR detector.
Before calculating various moments of net-proton distributions for
one centrality, such as $0-5\%$, $5-10\%$ , we should consider the
so called {\it Centrality Bin Width Effect} (CBWE) arising from the
impact parameter fluctuations due to the finite centrality bin
width. This effect must be corrected, as it may cause different
centrality dependence. To formulate and demonstrate the centrality
bin width effect, we write the event-by-event net-proton
distributions in one centrality:
\begin{eqnarray}
P(N) = \sum\limits_i {\omega _i } f^{(i)} (N), (\sum\limits_i
{\omega _i }=1),
\end{eqnarray} where the $\omega_{i}$ and $f^{(i)} (N)$ are the weighted and
net-proton distributions for $i$th impact parameter in one
centrality, respectively. From eqs. (2)-(5) and (11), we can
calculate the various order cumulants for the distribution $P(N)$ as
below:
\begin{eqnarray}
C_{1,N}  &=& \sum\limits_i^{} {\omega _i } C_{1,N}^i {\rm{ }}=\sum\limits_i^{} {\omega _i }<N>_i {\rm{ }}   \\
C_{2,N} &=& (\sum\limits_i^{} {\omega _i } C_{2,N}^i ) +
C^{'}_{2,C_{1,N}^i  } \\
C_{3,N}  &=& {\rm{(}}\sum\limits_i^{} {\omega _i } C_{3,N}^i ) +
C^{'}_{3,C_{1,N}^i }+3\times C^{'}_{{\rm{1,}}C_{1,N}^i {\rm{,1,}}C_{2,N}^i } \\
C_{{\rm{4}},N}  &=& (\sum\limits_i^{} {\omega _i } C_{4,N}^i {\rm{)
+
}}C^{'}_{4,C_{1,N}^i }+4\times C^{'}_{1,C_{1,N}^i ,1,C_{3,N}^i }  \nonumber \\
&+&6\times C^{'}_{1,(C_{1,N}^i )^2 ,1,C_{2,N}^i }-12\times
(C^{'}_{1,C_{1,N}^i })(C^{'}_{1,C_{1,N}^i ,1,C_{2,N}^i }) \nonumber \\
&-& 3\times (C^{'}_{2,C_{1,N}^i } )^2+3\times C^{'}_{2,C_{2,N}^i },
\end{eqnarray} where the $C_{k,N}^i$ ($k=1,2,3,4$) are the $k^{th}$ order
cumulant for net-proton distribution $f^{(i)} (N)$; the
$C^{'}_{k,X}$ ($X=C_{m,N}^i$, $k,m=1,2,3,4$) are $k^{th}$ order
cutmulant for random variable $X=C_{m,N}^i$ under the discrete
probability distribution $Prob(X)$=$\omega_i$; the
$C^{'}_{1,X,1,Y}$=$<XY>-<X><Y>$ ($X=C_{k,N}^i$, $k=1,2,3,4$) are the
first order joint cumulant for random variable $X,Y$ under the
discrete probability distribution $Prob(X,Y)$=$\omega_i$. We find
that the higher order cumulants $C_{k,N}$ ($k=1,2,3,4$) can be
expressed by the addition of two parts, one is the weighted average
of the same order cumulant of each sub-distribution  $ f^{(i)} (N)$,
and the other part is the cumulant of lower order cumulant under the
discrete weighted distributions, which stems from the fluctuation of
impact parameters within the centrality and results in the CBWE.

Experimentally, the smallest centrality bin is determined by a
single uncorrected reference multiplicity value measured by TPC.
Generally, we usually report our results for a wider centrality bin,
such as $0-5\%$,$5-10\%$,...etc., to supress the statistical
fluctuations. To eliminate the centrality bin width effect, we
calculate the various moment for each single $N_{ch}$ within one
wider centrality bin and weighted averaged by the number of events
in each $N_{ch}$.
\begin{eqnarray}
\sigma  &=& \frac{{\sum\limits_r^{} {n_r \sigma _r  }
}}{{\sum\limits_r^{} {n_r } }} = \sum\limits_r^{} {\omega _r \sigma
_r  },    \\
S &=& \frac{{\sum\limits_r^{} {n_r S_r } }}{{\sum\limits_r^{} {n_r }
}} = \sum\limits_r^{} {\omega _r } S_r, \\
\kappa  &= &\frac{{\sum\limits_r^{} {n_r \kappa _r }
}}{{\sum\limits_r^{} {n_r } }} = \sum\limits_r^{} {\omega _r }
\kappa _r,
\end{eqnarray} where the $n_r$ is the number of events in $r^{th}$ refmult and
the corresponding weight $\omega_r={{ {n_r  } }}/{{\sum\limits_r^{}
{n_r } }}$.

\section{Results}
In this section, we will present beam energy and system size
dependence for the various moments ($M,\sigma,S,\kappa$) as well as
moment products ({\KV},{\SD}) of net-proton distributions for Au+Au
collisions at {\sNN} = 19.6, 39, 62.4, 130, 200 {\gev}, Cu+Cu
collisions at {\sNN} = 22.4, 62.4, 200 {\gev}, d+Au collisions at
{\sNN} = 200 {\gev} and p+p collisions at {\sNN} = 62.4, 200 {\gev}.
First, we will show the typical event-by-event net-proton
multiplicity distributions from different colliding systems. Then we
studied the centrality as well as energy dependence of various
moments and moment products. The systematic errors are estimated by
varying the following requirements for $p(\bar{p})$ tracks: DCA,
track quality reflected by the number of ot points used in track
reconstruction and the $dE/dx$ selection criteria for $p(\bar{p})$
identification. The statistical and systematic error are shown
separately by lines and brackets, respectively.

\subsection{Event-by-Event Net-proton Multiplicity Distributions}
\begin{figure}[htb]
\begin{center}
\includegraphics[width=0.5\textwidth]{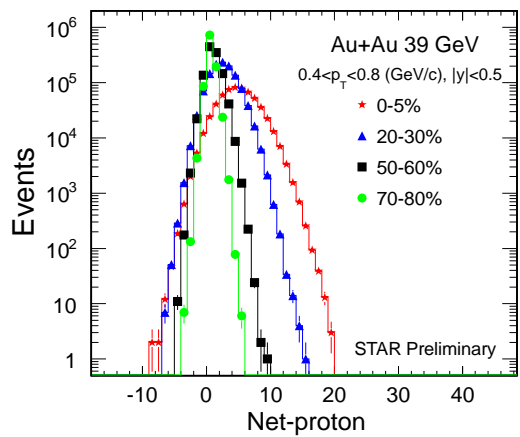}
\caption[Net-proton distributions for Au+Au central
collisions]{Typical event-by-event net-proton multiplicity
distributions for Au+Au collisions at {\sNN} = 39 {\gev}.}
\label{fig:netproton_dis_AuAu}
\end{center}\end{figure}

Event-by-event net-proton multiplicity distributions for Au+Au
collisions at {\sNN} = 39 GeV measured within $0.4<p_{T}<0.8$
{\gevc} and $|y|<0.5$ are shown in Fig.
\ref{fig:netproton_dis_AuAu}. Going from peripheral to central
collisions, it is found that the distributions become wider and more
symmetric for central collisions.

\subsection{Centrality Dependence of Moments and Moment Products of Net-proton Distributions}
\begin{figure}[htb]
 \hspace{-2cm}
\begin{minipage}[t]{0.6\linewidth}
\centering \vspace{0pt}
    \includegraphics[scale=0.45]{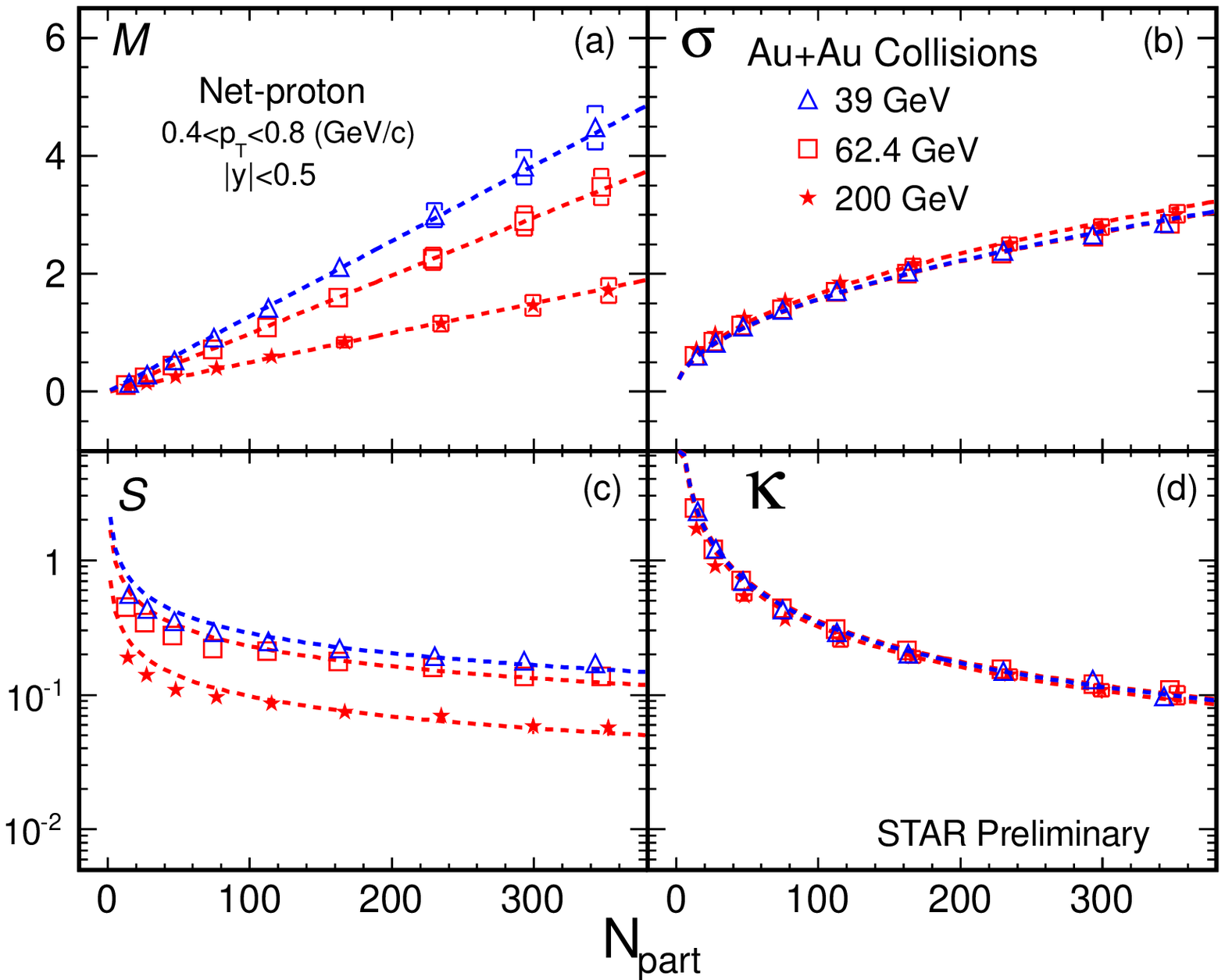}
   \caption{Centrality dependence of various moments of net-proton
multiplicity distributions for Au+Au collisions at {\sNN} =39, 62.4,
200 {\gev}. The dashed lines shown in the figure are expectation
lines from Central Limit Theorem (CLT).}
\label{fig:scaling_AuAu_all}
  \end{minipage}%
  \hspace{0.1in}
  \begin{minipage}[t]{0.6\linewidth}
  \centering \vspace{0pt}
   \includegraphics[scale=0.4]{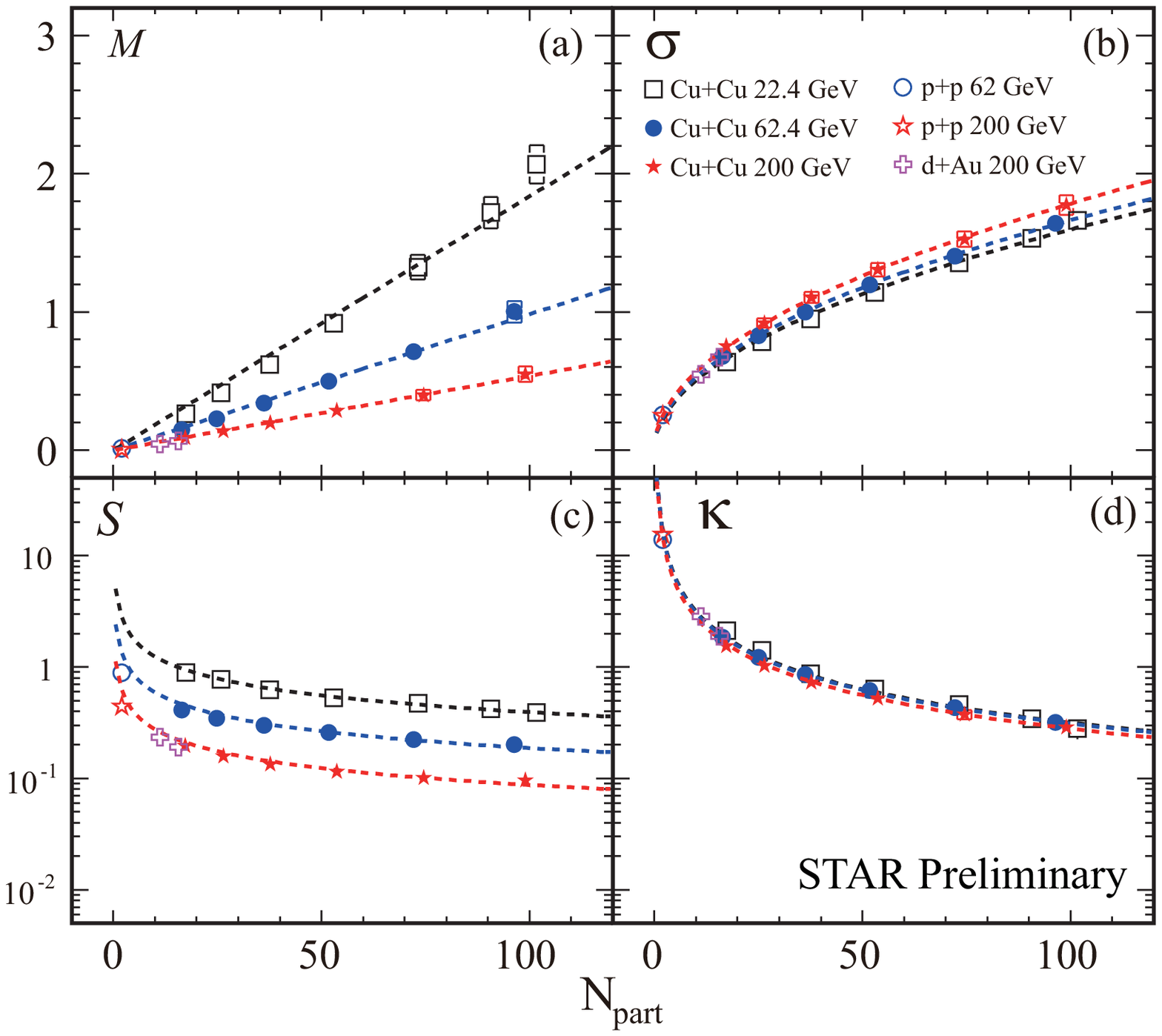}
    \caption[{\KV} of net-proton distributions as a function of Nsigma proton ( $Z_{p}$ )]
    {Centrality dependence of various moments of
net-proton multiplicity distributions for Cu+Cu collisions at {\sNN}
= 22.4, 62.4 and 200 {\gev}, d+Au collisions at {\sNN} = 200 {\gev}
and p+p collisions at {\sNN} = 62.4 and 200 {\gev}. The dashed lines
shown in the figure are expectation lines from Central Limit Theorem
(CLT).} \label{fig:scaling_CuCu_pp}
  \end{minipage} %
\end{figure}

The centrality ($N_{part}$) dependence for various moments ($M,
\sigma, S, \kappa$) of net-proton multiplicity distributions from
Au+Au collisions at {\sNN}=39, 62.4 and 200 GeV, Cu+Cu collisions at
{\sNN}=22.4, 62.4 and 200 GeV, d+Au collisions at {\sNN}=200 GeV and
p+p collisions at {\sNN}=62.4 and 200 GeV are shown in Fig.
\ref{fig:scaling_AuAu_all} and Fig. \ref{fig:scaling_CuCu_pp}. We
find that $M$ and $\sigma$ increase with $N_{part}$ monotonically,
while $S$ and $\kappa$ decrease with $N_{part}$. The centrality and
energy dependence of $S$ and $\kappa$ indicate that the net-proton
distributions become more symmetric for more central collision and
higher energies.

The dashed lines in the Fig. \ref{fig:scaling_AuAu_all} and Fig.
\ref{fig:scaling_CuCu_pp} represent the expectations from Central
Limit Theorem (CLT) when assuming the superposition of many
identical and independent particle emission sources in the
system~\cite{PRL,SQM2009}. In Fig. \ref{fig:scaling_AuAu_all} and
\ref{fig:scaling_CuCu_pp}, the centrality dependence of various
moments can be well described by the dashed lines expected from CLT.
Especially in Fig. \ref{fig:scaling_CuCu_pp}, the various moments of
p+p and d+Au collisions follow the CLT lines of Cu+Cu collision at
the corresponding energy very well. This also supports the identical
independent emission sources assumption.
\begin{figure}[htb]
\begin{center}
\includegraphics[width=0.6\textwidth]{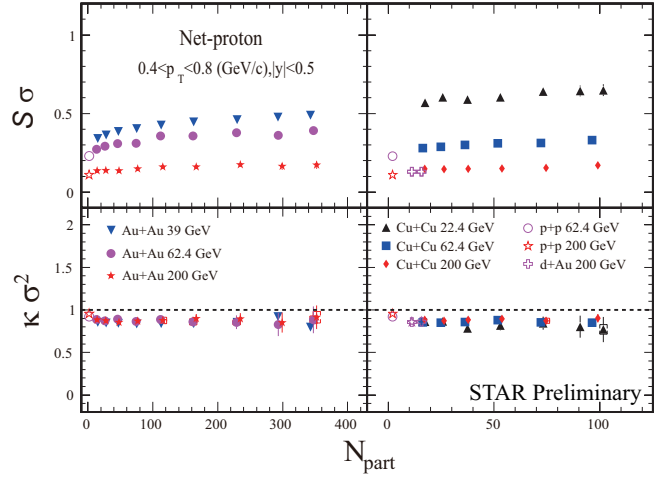}
\caption[{\KV} and {\SD} of net-proton distributions ]{Moment
products ({\KV} and {\SD}) of net-proton distributions for Au+Au,
Cu+Cu, d+Au and p+p collisions.} \label{fig:Moment Products1}
\end{center}\end{figure}
Fig. \ref{fig:Moment Products1} shows the centrality dependence of
moment products $S \sigma$ and $\kappa \sigma^2$ of net-proton
distributions, which are directly related to the baryon number
susceptibility ratio in Lattice QCD and HRG models as $\kappa
\sigma^2=\chi^{(4)}_{B}$/$\chi^{(2)}_{B}$ and $S
\sigma=\chi^{(3)}_{B}$/$\chi^{(2)}_{B}$, for p+p, Cu+Cu and Au+Au
collisions at various colliding energies. $S \sigma$ shows a weak
increase with centrality, while the $\kappa \sigma^2$ shows no
centrality dependence.

\subsection {Energy Dependence of the Moment Products ({\SD} and {\KV})}

\begin{figure}[htb]
 \hspace{-2cm}
\begin{minipage}[t]{0.6\linewidth}
\centering
\vspace{0pt}
\includegraphics[width=0.8\textwidth]{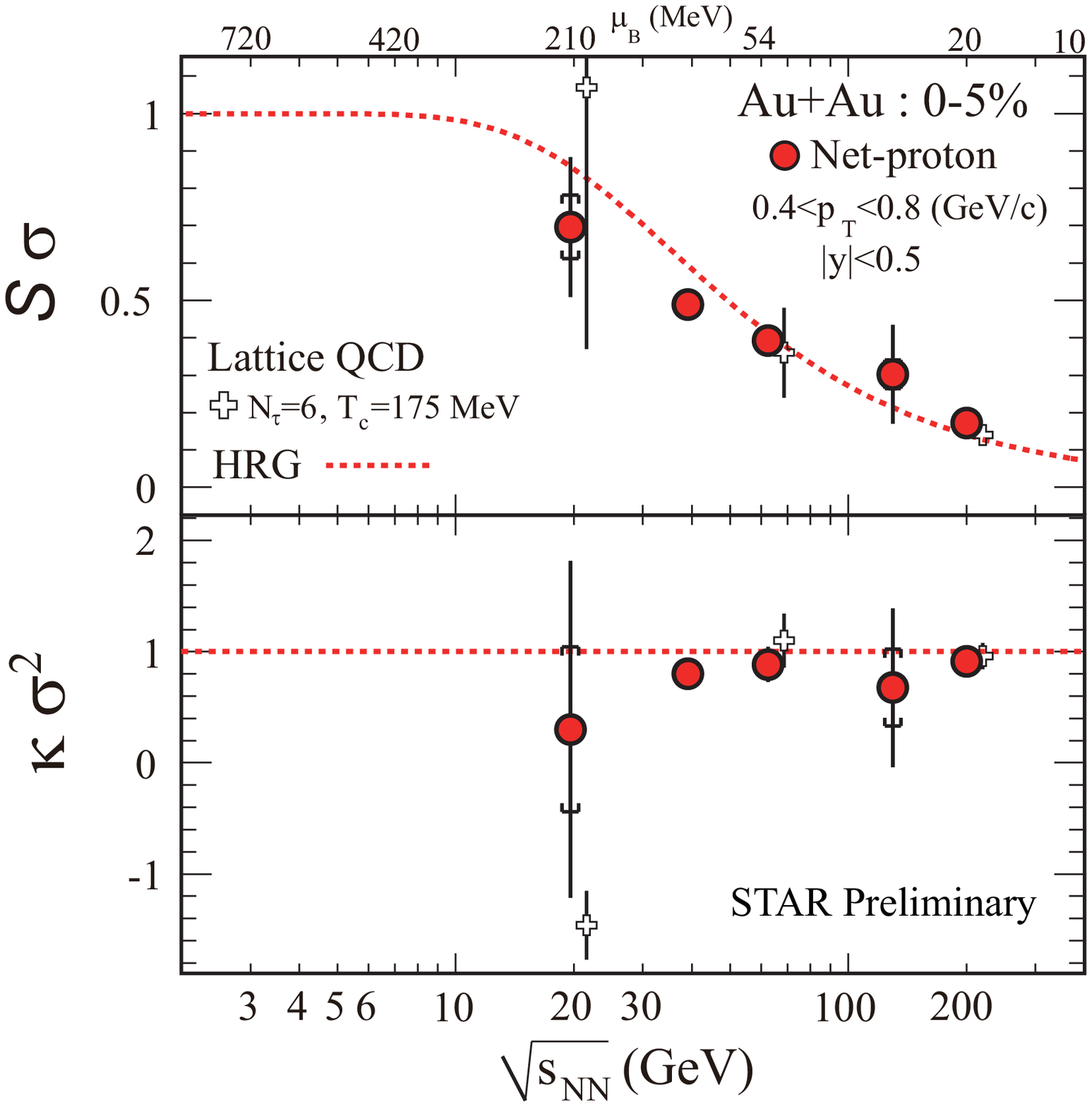}
\caption[Energy dependence of {\KV} and {\SD} for Au+Au
collisions]{Energy dependence of moment products ({\KV} and {\SD})
of net-proton distributions for central Au+Au collisions ($0-5\%$,
19.6 GeV: $0-10\%$, 130 GeV: $0-6\%$). The red dashed lines denote
the HRG model calculations, in which the $S\sigma=\tanh(\mu_{B}/T)$
and {\KV}=1. The empty markers denote the results calculated from
Lattice QCD~\cite{science}. } \label{fig:Moment Products}
  \end{minipage}%
  \hspace{0.1in}
  \begin{minipage}[t]{0.6\linewidth}
  \centering \vspace{0pt}
  \includegraphics[width=0.8\textwidth]{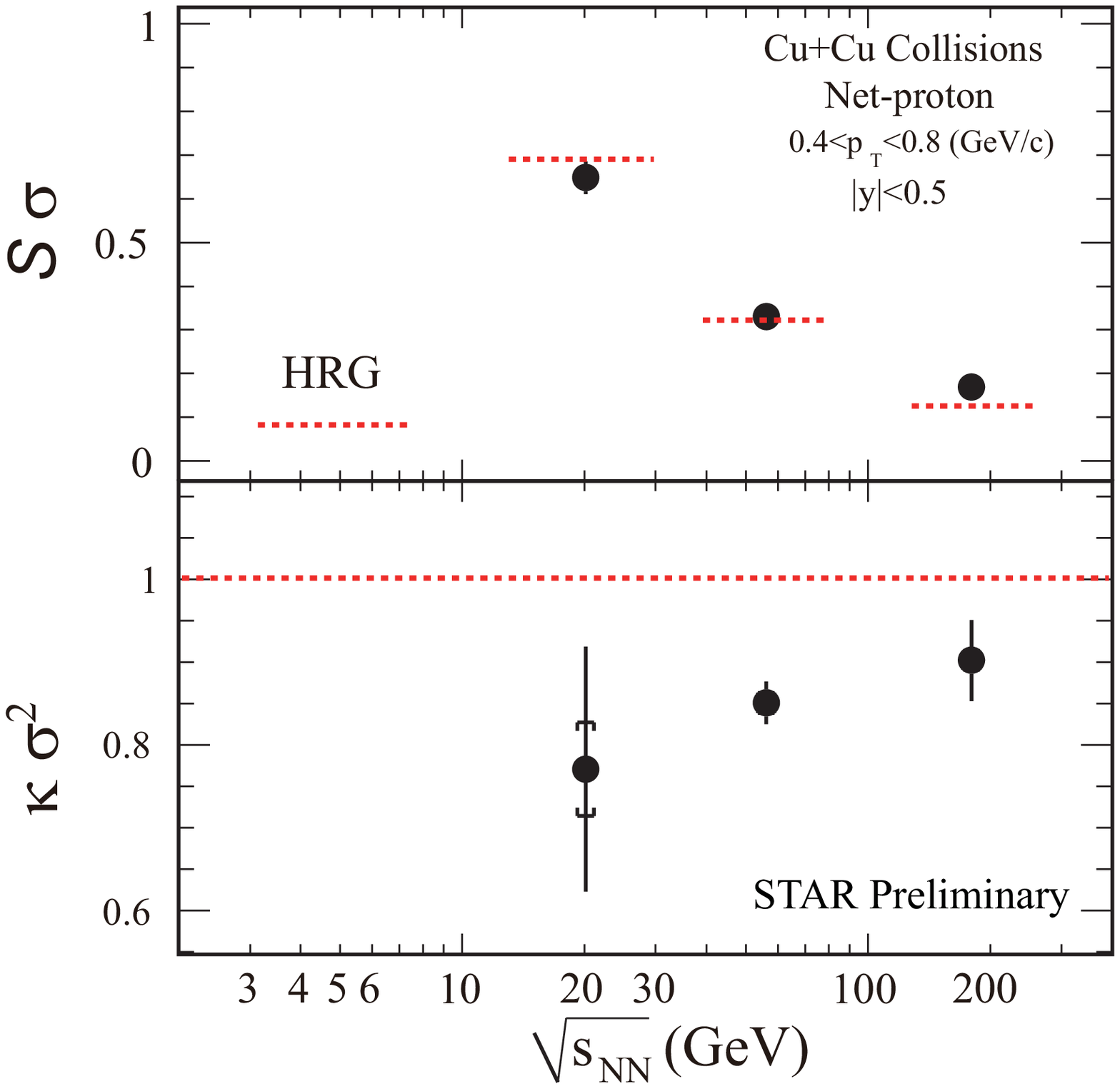}
\caption[Energy dependence of moment products ({\KV},{\SD}) for
Cu+Cu central collisions.]{Energy dependence of moment products
({\KV} and {\SD}) of net-proton distributions for Cu+Cu central
collisions ($0-10\%$, 22.4 GeV: $0-5\%$). The dashed lines shown in
the figures are from HRG model calculations, in which $S
\sigma=\tanh(\mu_{B}/T)$ and {\KV}=1. The $\mu_{B}$ and $T$ values
are taken from~\cite{Scaling}.} \label{fig:SD_KV_CuCu}
  \end{minipage} %
\end{figure}

In Fig. \ref{fig:Moment Products}, we show the energy dependence of
the $S \sigma$ and $\kappa \sigma^2$ of net-proton distributions for
most central Au+Au collisions ($0-5\%$, 19.6 GeV: $0-10\%$, 130 GeV:
$0-6\%$). Lattice QCD~\cite{science} and HRG model~\cite{HRG}
calculations are also shown for comparison. Lattice QCD results are
obtained with time extent $N_{\tau}=6$ and phase transition
temperature at $\mu_B$=0, $T_{c}=175$ MeV. The red dashed lines of
the HRG model in the upper panel and lower panels are evaluated by
$S \sigma=\tanh(\mu_{B}/T)$ and $\kappa \sigma^{2}=1$, respectively,
where the $\mu_{B}/T$ ratio at chemical freeze-out is parameterized
as a function of colliding energy based on
reference~\cite{Chemical}. The corresponding baryon chemical
potential ($\mu_B$) at chemical freeze-out for each energy is shown
in the upper band of the Fig. \ref{fig:Moment Products}. We find
that the moment products ({\KV} and {\SD}) of Au+Au collisions at
{\sNN}=200, 130, 62.4 {\gev} are consistent with Lattice QCD and HRG
model calculations. The $S \sigma$ and $\kappa \sigma^2$ for Au+Au
collisions at {\sNN} = 39 {\gev} deviates from HRG model
calculations. Surprisingly, $\kappa \sigma^2$ from Lattice QCD
calculations at {\sNN} = 19.6 {\gev} show a negative
value~\cite{science}. However, due to the limited statistics, the
statistical errors of experimental data are large at 19.6 GeV. The
STAR is running to take more data at 19.6 GeV in year 2011. Those
deviations could be linked to the chiral phase
transition~\cite{chiral_HRG} and presence of QCD critical
point~\cite{Neg_Kurtosis}. Recent linear $\sigma$ model calculations
demonstrate that the forth order cumulant of the fluctuations for
$\sigma$ field will be universally negative, when the QCD critical
point is approached from cross-over side~\cite{Neg_Kurtosis}. It
will cause the measured {\KV} as well as kurtosis ($\kappa$) of
net-proton distributions to be smaller than their Poisson
expectation values.

Fig. \ref{fig:SD_KV_CuCu} shows the energy dependence of {\KV} and
{\SD} for Cu+Cu central collisions. The red dashed lines in the
figure are obtained from the HRG model by using the formula $S
\sigma=\tanh(\mu_{B}/T)$, where the $\mu_B$ and $T$ are from thermal
model fits of the particle ratios. We find that our experimental
data is consistent with HRG model expectations for {\SD} of
net-proton distributions. While the {\KV} deviates from HRG model
calculations and monotonically decrease as the collision energy
decreases.

\subsection{Charged Particle Density ({\dNdeta}) Scaling of {\SD} and Evidence of Thermalization in Heavy Ion Collisions}

\begin{figure}[htb]
 \hspace{-2cm}
\begin{minipage}[t]{0.6\linewidth}
\centering \vspace{0pt}
\includegraphics[width=0.8\textwidth]{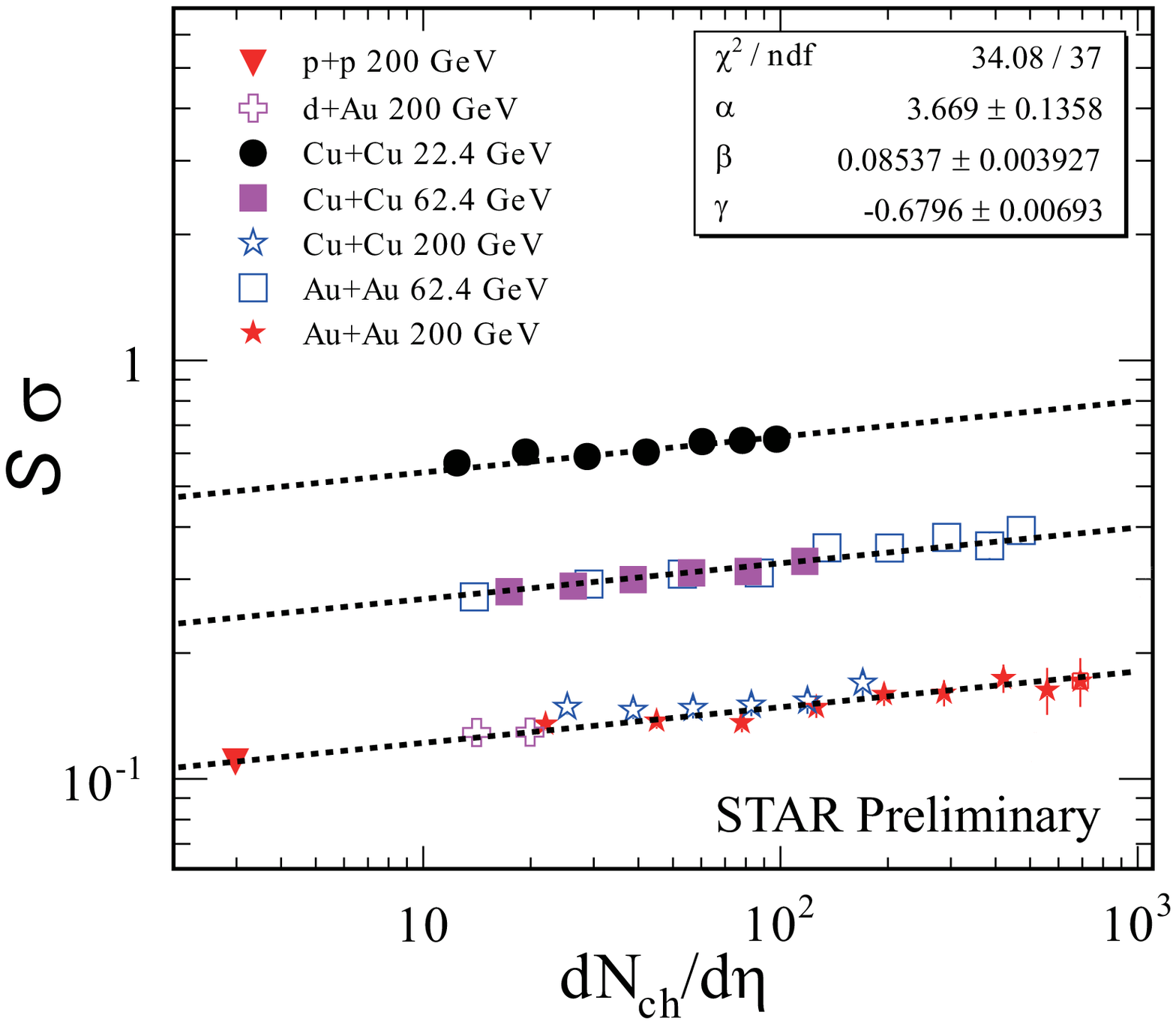}
\caption{{\SD} of net-proton distributions as a function of charged
particle density at mid-rapidity ({\dNdeta}) for various colliding
systems. The dashed lines in the figures is the fitting lines.}
\label{fig:SD_scaling}
  \end{minipage}%
  \hspace{0.1in}
  \begin{minipage}[t]{0.6\linewidth}
  \centering \vspace{0pt}
\includegraphics[width=0.8\textwidth]{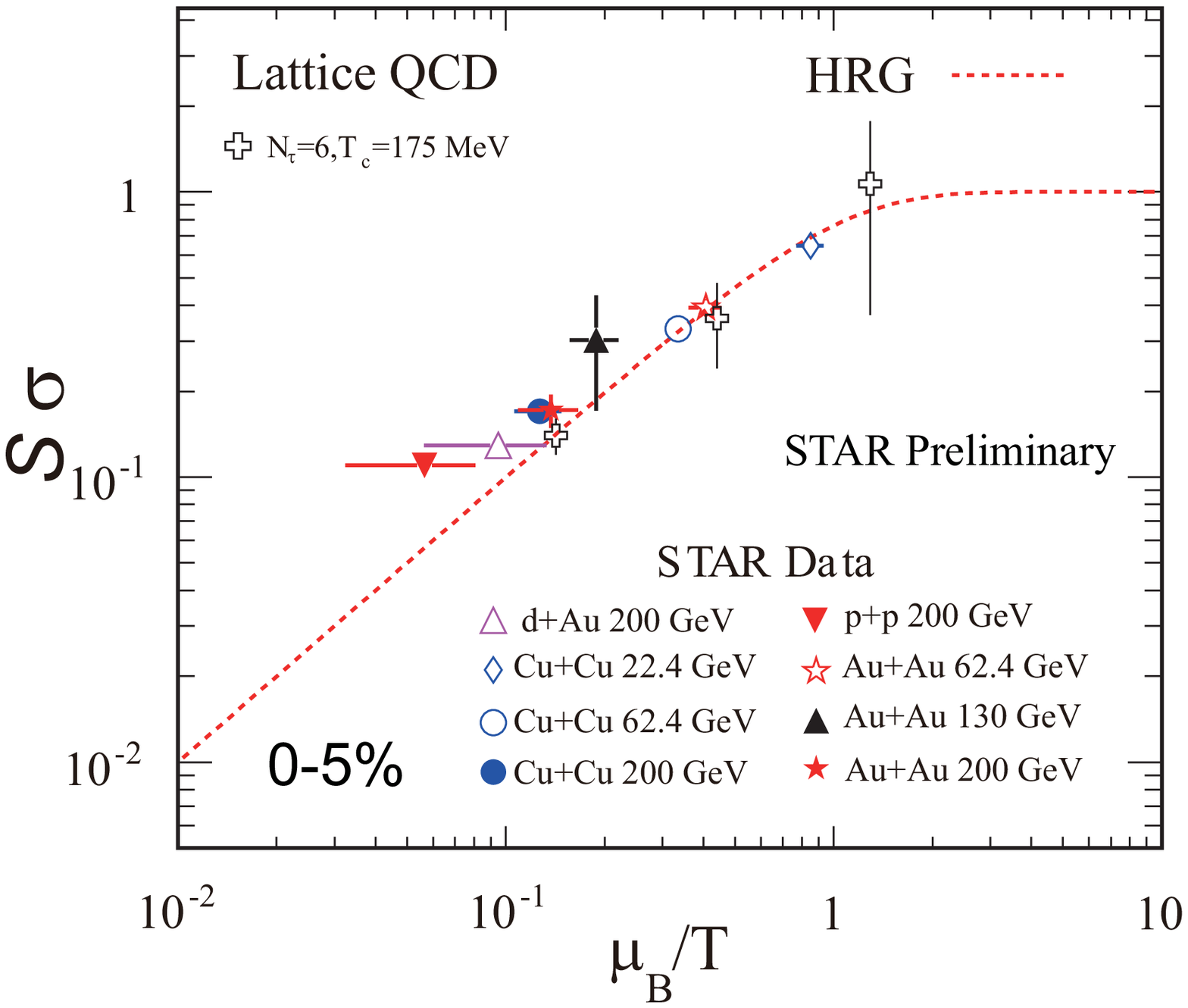} \caption[{\SD}
versus baryon chemical potential over temperature ratio
($\mu_{B}/T$)]{{\SD} of net-proton distributions as a function of
baryon chemical potential over temperature ratio  ($\mu_{B}/T$) for
most central collisions ($0-5\%$, 130 GeV: $0-6\%$, d+Au :
$0-20\%$). The red dashed line in the figures is the result of the
HRG model. Lattice QCD results with $N_{\tau}=6$ and $T_{c}=175$ MeV
are also shown in the figure.} \label{fig:SD_muT}
  \end{minipage} %
\end{figure}

The $S \sigma$ of net-proton distributions for various colliding
systems including Au+Au collisions at {\sNN} = 62.4 and 200 GeV,
Cu+Cu collisions at {\sNN} = 22.4, 62.4 and 200 GeV, d+Au and p+p
collisions at {\sNN} = 200 GeV, as a function of $dN_{ch}/d\eta$ are
shown in the Fig. \ref{fig:SD_scaling} with double logarithm axis.
It is obvious that for a fixed colliding energy, such as {\sNN}=200
GeV, the moment products of $S \sigma$ of net-proton distributions
for different system size the p+p, d+Au, Cu+Cu and Au+Au collisions
have power law dependence on the charged particle density
($dN_{ch}/d\eta$). Thus, we fit the $S \sigma$ of net-proton
distributions for various colliding systems with double power law
formula: \be S
\sigma(dN_{ch}/d\eta,\sqrt{s_{\mathrm{NN}}})=\alpha\times(dN_{ch}/d\eta)^{\beta}\times(\sqrt{s_{\mathrm{NN}}})^{\gamma}
\ee The fitting results are shown in the Table. \ref{tab:SD_fit}.
\begin{table}[htbp] \centering
\caption[Fitting parameters for {\SD} as a function of
{\dNdeta}]{Fitting parameters for {\SD} as a function of {\dNdeta}}
\label{tab:SD_fit} \vspace{0.03\textwidth} \centering
\begin{tabular}{ccc}
\hline Parameters & Value & Approx.\tabularnewline \hline \hline
$\chi^{2}/ndf$ & 34.08/37 & $0.92$\tabularnewline \hline $\alpha$ &
3.669$\pm0.1358$ & $\frac{11}{3}$\tabularnewline \hline $\beta$ &
0.0853$\pm0.003927$ & $\frac{1}{12}$\tabularnewline \hline $\gamma$
& -0.6796$\pm0.00693$ & $-\frac{2}{3}$\tabularnewline \hline
\end{tabular}
\end{table}
The $S \sigma$ can be well described by the power law formula $S
\sigma=\frac{11}{3}\times(\frac{1}{s^{4}}\frac{dN_{ch}}{d\eta})^{\frac{1}{12}}$,
where the $s$ is the square of the center of mass energy. For high
energy heavy ion collisions the temperature is approximately
constant and the ratio $\mu_{B}/T<<1$, thus we have the
approximation $\mu_{B}/T\sim \tanh(\mu_{B}/T)=S
\sigma=\frac{11}{3}\times(\frac{1}{s^{4}}\frac{dN_{ch}}{d\eta})^{\frac{1}{12}}$.
This denotes the relation between $\mu_{B}/T$ and the charged
particle density and colliding energy for high energy nuclear
collisions.

Multiplicity fluctuations and inclusive yields are two basic
properties in high energy heavy ion collisions. For a thermal
system, both the fluctuations and yields should be described by the
thermodynamic parameters ($\mu_B$ and $T$), which completely
determine the properties of the thermal system. The fluctuation
observable $S \sigma$ of most central net-proton distributions
versus thermodynamic parameter $\mu_{B}/T$ ratios, which are
extracted from the thermal model fit of the particle ratio, is shown
in the Fig. \ref{fig:SD_muT} for various colliding systems. Lattice
QCD calculations with $N_{\tau}=6$ and $T_{c}=175$ MeV and the HRG
model relation  $S \sigma=\tanh(\mu_{B}/T)$ are also plotted in Fig.
\ref{fig:SD_muT} for comparison. We find that high energy heavy ion
collisions, such as Au+Au and Cu+Cu collisions, are consistent with
Lattice QCD and HRG model calculations, while the elementary p+p
collision deviate from the HRG model calculations. In addition to
the perfect description of the particle yields by the thermal model,
the agreement of higher order fluctuations with thermal model
predications provide further evidence that the colliding system has
achieved thermalization in most central high energy heavy ion
collision.

\section{Summary}
The higher moments of net-proton multiplicity distributions measured
in heavy ion collision experiment have been applied to search for
the QCD critical point, due to the high sensitivity to the
correlation length ($\xi$). The beam energy and system size
dependence for higher moments ($M, \sigma, S, \kappa$) as well as
moment products ({\KV, \SD}) of net-proton multiplicity
distributions have been presented with a broad energy range and
different system sizes, which include Au+Au collisions at {\sNN} =
200, 130, 62.4, 39 and 19.6 GeV, Cu+Cu collisions at {\sNN} = 200,
62.4 and 22.4 GeV, d+Au collisions at {\sNN} = 200 GeV, p+p
collisions at {\sNN} = 200 and 62.4 GeV. The moment product {\KV}
shows no centrality dependence while \SD\ shows a weak centrality
dependence. The energy dependence is studied by comparing the
results from the Au+Au 200 GeV to those from the BES energies. The
moment products {\KV} and {\SD} of net-proton distributions from
most central Au+Au collisions are consistent with Lattice QCD and
HRG model calculations at high energy (200, 130, 62.4 GeV) while
deviating from (smaller than) HRG model calculations at \sNN\ = 39
GeV. Lattice QCD calculations show negative value for \KV\ at \sNN\
= 19.6 GeV. The deviations could potentially be linked to chiral
phase transitions and QCD critical point. But the experimental data
is with large error bar due to the limited statistics. Fortunately,
this ambiguity can be clarified soon by 19.6 GeV data taken in year
2011 with higher statistics. Recent model calculations show that the
{\KV} value will always be smaller than its Poisson statistical
value 1, when QCD critical point is approached from the high energy
cross-over side.

On the other hand, the mutual agreements between the $\mu_B/T$
extracted from thermal model fits of particle ratio and from the
event-by-event fluctuations observable {\SD} of net-proton
distributions provides further evidence of thermalization of the hot
dense matter created in the heavy ion collisions. Further, the {\KV}
and {\SD} of net-proton distributions are consistent with Lattice
QCD and HRG model calculations at high energy, which support the
thermalization of the colliding system. The deviations of the {\KV}
and {\SD} of net-proton distributions for Au+Au central collisions
from HRG model predications at low energies are not well understood.
This may result from the non-applicability of grand canonical
ensemble or the appearance of QCD critical point and chiral phase
transitions at low energies. However, it should be further
investigated.

\section*{Acknowledgement}
We thank S. Gupta, F. Karsch, K. Rajagopal, K. Redlich, M. Stephanov
for enlightening discussions. We thank the RHIC Operations Group and
RCF at BNL, the NERSC Center at LBNL and the Open Science Grid
consortium for providing resources and support. This work was
supported in part by the Offices of NP and HEP within the U.S. DOE
Office of Science, the U.S. NSF, the Sloan Foundation, the DFG
cluster of excellence `Origin and Structure of the Universe' of
Germany, CNRS/IN2P3, FAPESP CNPq of Brazil, Ministry of Ed. and Sci.
of the Russian Federation, NNSFC, CAS, MoST, and MoE of China, GA
and MSMT of the Czech Republic, FOM and NWO of the Netherlands, DAE,
DST, and CSIR of India, Polish Ministry of Sci. and Higher Ed.,
Korea Research Foundation, Ministry of Sci., Ed. and Sports of the
Rep. Of Croatia, and RosAtom of Russia.

\section*{References}
\bibliography{WWND2011}
\bibliographystyle{unsrt}

\end{document}